\documentclass[journal]{IEEEtran}
\usepackage{amsthm, amssymb, amsmath, color,amsfonts}
\usepackage{graphicx}
\usepackage{optidef}
\usepackage[update,prepend]{epstopdf}
\usepackage[noadjust]{cite}
\usepackage[latin1]{inputenc}
\usepackage{tikz}
\usetikzlibrary{arrows,calc}		
\usepackage{bbm} 
\usepackage{bm}
\usepackage{pdfpages}
\usepackage{tabularx}
\usepackage{multirow}
\usepackage{subfigure}
\usepackage{comment}
\usepackage{algorithm}
\usepackage{algpseudocode}
\DeclareGraphicsExtensions{.eps}

\def\nb0{{\mathbf{0}}}
\def\nb1{{\mathbf{1}}}

\newtheorem{lemma}{Lemma}

\newtheorem{theorem}{Theorem}
\newtheorem{prop}{Proposition}

\def\figref#1{Fig.\,\ref{#1}}%

\allowdisplaybreaks 

\begin{document}
\graphicspath{{./Figures/}}
\title{
Determinantal Learning for Subset Selection in Wireless Networks
}
\author{
Xiangliu Tu, Chiranjib Saha,~\IEEEmembership{Member,~IEEE}, Harpreet S. Dhillon,~\IEEEmembership{Fellow,~IEEE}

\thanks{X. Tu and H. S. Dhillon are with Wireless@VT, Bradley Department of Electrical and Computer Engineering, Virginia Tech, Blacksburg, VA, 24061, USA. Email: \{xiangliutu, hdhillon\}@vt.edu. C. Saha is with the
Qualcomm Standards and Industry Organization, Qualcomm Technologies
Inc., San Diego, CA 92121 USA. Email: csaha@qti.qualcomm.com. The support of the US NSF (Grant CNS-1923807) is gratefully acknowledged. Some preliminary results from this article were presented at~\cite{saha2019machine} and~\cite{tu2023determinantal}. } 
}

\maketitle
\begin{abstract}
Subset selection is central to many wireless communication problems, including link scheduling, power allocation, and spectrum management.
However, these problems are often NP-complete, because of which heuristic algorithms applied to solve these problems struggle with scalability in large-scale settings.
To address this, we propose a determinantal point process-based learning (DPPL) framework for efficiently solving general subset selection problems in massive networks. 
The key idea is to model the optimal subset as a realization of a determinantal point process (DPP), which balances the trade-off between quality (signal strength) and similarity (mutual interference) by enforcing negative correlation in the selection of {\em similar} links (those that create significant mutual interference).
However, conventional methods for constructing similarity matrices in DPP impose decomposability and symmetry constraints that often do not hold in practice. 
To overcome this, we introduce a new method based on the Gershgorin Circle Theorem for constructing valid similarity matrices.
The effectiveness of the proposed approach is demonstrated by applying it to two canonical wireless network settings: an ad hoc network in 2D and a cellular network serving drones in 3D. 
Simulation results show that DPPL selects near-optimal subsets that maximize network sum-rate while significantly reducing computational complexity compared to traditional optimization methods, demonstrating its scalability for large-scale networks.

\end{abstract}


\section{Introduction} \label{sec:intro}

An important class of resource management problems in wireless networks, such as power control, link scheduling, network utility maximization, and beamformer design requires subset selection.
The goal of these problems is to determine an optimal subset from the ground set according to specific objective functions. 
Typically, heuristic algorithms are designed to find a local optimum with acceptable complexity by using various optimization tools, such as geometric programming (GP), integer linear or non-linear programming~\cite{cooper1981survey,burer2012non,chiang2005geometric,chiang2007power}.  
However, since subset selection problems are often NP-complete, solving them efficiently with these heuristic algorithms gets challenging as the network size grows.
This is particularly problematic for solving resource allocation problems in massive Internet of Things (IoT) networks~\cite{dhillon7842431}. Not surprisingly, problems such as those related to interference management in IoT networks are of interest to both industry~\cite{3GPP123456} and academia~\cite{Na8565904,li2019interference}. 
Therefore, to overcome the scalability challenge, we turn our attention to the determinantal point process (DPP), a tool grounded in stochastic geometry (SG) that has sound applications in popular machine learning (ML) problems, such as recommender systems and document summarization~\cite{wilhelm2018practical,cho2019improving}. 
The key idea is to view the optimal subset as a realization of a DPP, in which items with high quality and low similarity (with each other) are preferentially selected from the ground set. 
This approach reduces the subset selection problem to sampling from a DPP whose parameters are trained for a given subset selection problem.

Within the wireless community, DPPs have been successfully applied to model and analyze cellular networks~\cite{li2015statistical,miyoshi2014cellular}. 
The conference version of this paper~\cite{saha2019machine} was the first to introduce a finite DPP-based learning (DPPL) framework for efficiently solving link scheduling problems in 2D ad hoc networks. 
A notable advantage of the DPPL framework is its ability to capture diversity among items through an appropriately defined similarity model. 
However, extending this framework to new settings presents a significant challenge: the similarity matrices must be positive semidefinite (PSD). 
Conventional methods for constructing these matrices, such as cosine similarity or covariance function, guarantee PSD property at the expense of inducing additional decomposability and symmetry constraints on them, which might not always be satisfied in practice~\cite{lavancier2015determinantal}. 
For instance, in cellular networks where base stations (BSs) are equipped with directional antennas, modeling interference as the similarity between user-BS links violates these constraints.
To address this limitation, we provide a new method using the Gershgorin Circle Theorem to construct valid similarity matrices. 
This approach enables the DPPL framework to handle the specific challenges of directional antennas, thereby capturing more complicated correlation structures in practical wireless networks. 

\subsection{Prior Art}  \label{sec:related}              

Because of their very nature, many subset selection problems, such as the ones appearing in link scheduling, power allocation, and spectrum management, are typically solved using heuristic algorithms. 
For instance, integer programming-based algorithms have been developed for handling subset selection problems with non-convex non-linear objective functions~\cite{cooper1981survey,burer2012non}.  
Additionally, GP has been extensively used to solve weighted sum-rate maximization problems, including the max weighted link scheduling in multihop wireless networks, link activations in multiple-input multiple-output (MIMO) networks, and power/rate allocation in wireless networks~\cite{weeraddana2012weighted,chiang2007power,elbatt2004joint,1634905, chiang2005geometric}. 
However, due to their complexity, most of these heuristic approaches face significant implementation challenges in large-scale networks.
As mentioned above, this challenge is particularly prominent in massive IoT networks.
Therefore, alternative approaches have long been desirable. 
As we will discuss shortly, DPPs offer one such alternative by treating the optimal solution as a realization of a point process.  

DPPs have found applications in various ML problems, such as classification~\cite{xie2017deep}, document summarization~\cite{cho2019improving}, and recommondation~\cite{wilhelm2018practical}. 
In these scenarios, DPPs efficiently capture the balance between the quality of items and the similarity among them.  
For example, a DPP-based recommendation system is designed to provide content that matches user preferences (quality) while ensuring that the content is not monotonous or repetitive (similarity). 
The quality and similarity of items can be jointly modeled by the kernel matrix $K$ or separately by {\it L-ensemble} in DPPs. 
The work of~\cite{Kulesza_2012,kulesza2012learning} proposes a machine learning framework that incorporates DPPs, which highlights the ability of parameterized DPPs to learn the trade-off between quality and similarity. 
Since the probability assigned by DPP is proportional to the determinant of the submatrix of $L$, it requires the matrix $L$ to be PSD~\cite{borodin2005eynard}.  
Although conventional approaches, such as the covariance function and the cosine similarity, are widely used to construct matrix $L$, these methods impose the triangle inequality property onto the similarity matrix, a constraint that may not always be satisfied in practice (as evident from one of the examples of this paper). 
This may unnecessarily limit the applicability of DPPs to certain real-world setups. 
Although the work of~\cite{gartrell2019learning} proposes skew-symmetric matrices to overcome these limitations, this approach remains unsuitable in many practical scenarios, such as when using interference between links to define similarity matrices in wireless networks with directional antennas.

In the context of wireless networks, DPPs have been found use in the SG-based modeling and analysis of cellular networks. 
These models leverage DPPs to capture spatial repulsion in BS locations, which cannot be captured by the more popular Poisson point process (PPP)~\cite{li2015statistical,miyoshi2014cellular}. 
Despite the growing interest in data-driven learning within the ML community, finite DPPs and related learning frameworks have seen limited applications in wireless networks.
In the work of~\cite{blaszczyszyn2019determinantal}, authors demonstrate that DPPs can effectively replicate the properties of certain hard-core point processes used for wireless network modeling (such as the Mat\'{e}rn type-II process) in a finite window. 
In~\cite{saha2019interference}, the authors characterize interference distribution in a wireless local area network with carrier sense multiple access (CSMA) by first modeling the transmitter (Tx) locations as a PPP and then modeling a set of active transmitters (the result of CSMA) as a DPP. 
Building on these developments, more works have explored additional application scenarios. 
For example, the study of~\cite{liu2022learning} proposes a beamforming designer based on the DPPL for single-group multicast beamforming where the designer selects the subset of users with lower channel magnitudes and channel directions that are as orthogonal as possible. 
Furthermore, the work in~\cite{gu2021scaling} introduces heuristic clustering schemes based on the DPPL framework to construct the network with evenly distributed BS clusters, thereby addressing the scalability issue in large-scale 5G-based vehicular networks.
However, all these referenced works, which are built on~\cite{saha2019machine}, rely on the covariance function or cosine to construct their similarity matrices (as was the case in~\cite{saha2019machine}), both of which come with inherent constraints and limited applicability. 
In this journal extension of~\cite{saha2019machine}, we relax these key limitations related to the construction of similarity matrices, thereby resulting in a more powerful framework that is much more broadly applicable to wireless network problems. 

\subsection{Contributions} \label{sec:contributions}
This paper presents a comprehensive treatment of general subset selection problems in large-scale networks using the DPPL framework. 
In addition, we present a new approach for designing similarity matrices to overcome the key limitations of conventional methods for the DPPL framework. 
We further apply this framework to the link scheduling problem in wireless communications, a classical NP-hard subset selection problem. 
Simulation results demonstrate the effectiveness and scalability of the proposed DPPL framework with the newly designed similarity matrices. 
Our main contributions are detailed as follows:

 \begin{enumerate} 
    \item 
    We propose a DPP-based learning framework to tackle subset selection problems in large-scale networks.
    The key idea is to represent the optimal subset as a realization of a DPP, reducing the problem to sampling from a DPP whose parameters are trained for the given problem.
    The proposed DPPL framework learns the trade-off between quality and similarity by increasing the likelihood of selecting high-quality items while enforcing negative correlations among highly similar ones. 
    As discussed shortly, a key contribution of this work is interpreting this quality-similarity trade-off in the context of subset selection in wireless networks. 
    Compared to heuristic approaches, such as GP-based methods, DPPL demonstrates remarkable efficiency and scalability in solving subset selection problems in large-scale networks.
    \item  
    A key challenge in extending the DPPL framework to new settings is ensuring that the similarity matrix has all its principal minors non-negative, i.e., PSD matrices.
    While conventional approaches, such as cosine similarity and covariance function, are extensively used to generate PSD matrices, they impose additional constraints of symmetry and decomposability onto matrices, which are not always satisfied in practice.
    To overcome these limitations, we develop a new method using the Gershgorin Circle Theorem, which enables DPPLs to capture more complicated correlations among items. 
    \item 
    We apply the proposed DPPL framework to the link scheduling problems to demonstrate its efficacy and scalability in subset selection tasks.
    In particular, we focus on the sum-rate maximization problem in two canonical settings. 
    The first is an ad hoc network setting in 2D, in which we model the ground set by a random process. 
    The second is a cellular network setting serving drones in 3D, where the cellular BS locations are modeled as a deterministic hexagonal grid. 
    The goal is to identify the optimal subset of active links from the ground set. 
    Naturally, the links with higher signal-to-interference-and-noise ratio (SINR) are favored as they contribute more to the overall sum-rate (quality). 
    On the other hand, the active links should exhibit some degree of repulsion to avoid high mutual interference (similarity). 
    With this insight, the optimal subset of links can be modeled as a DPP over the ground set of a given network.
    Simulation results show that DPPL achieves a sum-rate comparable to the near-optimal solution obtained by heuristic approaches in scenarios where the optimization problem remains computationally feasible. 
    Moreover, as network size increases, traditional optimization algorithms become computationally intractable, whereas the DPPL framework can still solve these problems efficiently.  
    Notably, DPPL generalizes effectively, providing meaningful solutions in significantly larger settings despite being trained on smaller instances where heuristic algorithms can still generate training data. 
 \end{enumerate}

 The rest of this paper is organized as follows. Section~\ref{sec:DPP} provides a comprehensive introduction to DPP, including the {\it L-ensemble} definition and intuitive geometric explanation of its decomposition. 
Section~\ref{sec:case1} analyzes the link scheduling problem in canonical ad hoc network settings, focusing on sum-rate maximization problem. 
Then, in Section~\ref{sec:case2}, we extend the system setting to the hexagonal cellular networks with directional antennas in 3D space and present our new generative method for similarity matrices.
Section~\ref{sec:conclu} concludes this paper.     
 
 \section{Determinantal Point Process}\label{sec:DPP}

 \subsection{Definition of DPPs} \label{subsec:definitionDPP}

In this section, we will introduce the concept of DPP on finite sets.
For a more comprehensive and pedagogical treatment of this topic and extensive surveys of the prior work, interested readers may refer to~\cite{kulesza2012determinantal}.
Specifically, we focus on the discrete cases, where DPPs are the probability measures over all subsets of a finite ground set $\mathcal{Y}$. Consider a finite set $\mathcal{Y} = \left\{ 1, \dots, N\right\}$ consisting of $N$ discrete items. For any subset $A\subseteq \mathcal{Y}$, a DPP is defined as:
\begin{align}
	\mathcal{P}(A \subseteq \mathbf{Y}) = \det(K_A),
	\label{eq::DPP::def::k}
\end{align}
where $\mathbf{Y} \sim \mathcal{P}$ is a random subset of $\mathcal{Y}$. 
The DPP is parameterized by a $ N \times N$ kernel matrix $K$ indexed by the elements of $\mathcal{Y}$ and $K_A \equiv \left[ K_{ij} \right]_{i,j \in A}$ represents the submatrix of $K$ indexed by elements in $A $.
We define $ \det( K_\varnothing) = 1 $. 
It is straightforward to infer from definition~\eqref{eq::DPP::def::k} that all principal minors of matrix $K$ are nonnegative and do not exceed 1.  
To develop the DPP learning framework, we next introduce an alternative definition of DPPs using the {\it L-ensemble} formalism~\cite{borodin2005eynard}. 
This approach is particularly advantageous for learning purposes which we will explore in the sequel. 
In this approach, a DPP is defined in terms of a matrix $L$ index by $Y\subseteq \mathcal{Y}$ as follows
\begin{align}
	\mathcal{P}_L (\mathbf{Y} = Y) = \frac{\det(L_Y)}{\sum_{Y^{\prime} \in 2^{\mathcal{Y}}} \det (L_{Y^{\prime}})} = \frac{\det (L_Y)}{\det (L+{\rm I})},
	\label{eq::DPP::def::L}
\end{align}
where $L_Y = \left[L_{ij}\right]_{i,j \in Y}$ and ${\rm I}$ is a $N\times N$ identity matrix. 
The last step follows from the identity $\sum_{Y^{\prime} \in 2^{\mathcal{Y}}} \det (L_{Y^{\prime}}) = \det (L+{\rm I})$ (see~\cite[Theorem 2.1]{kulesza2012determinantal} for proof). 
Definition in~\eqref{eq::DPP::def::L} directly specifies the probabilities for all possible subsets of $\mathcal{Y}$. This requires that all principal minors of matrix $L$ be nonnegative, i.e., $L$ is a PSD matrix. 
As highlighted in the work of~\cite{borodin2005eynard}, a matrix $L$ can be defined by any matrix whose principal minors are nonnegative. 
Such matrices satisfying this property are termed $P_0$ matrices. 
Although the symmetric PSD matrices form only a subset of the $P_0$ matrices space, most research utilizes symmetric PSD matrices to define $L$~\cite{prussing1986principal}.
This preference is primarily driven by the property of symmetric PSD matrices which allows $L$ to be expressed in the form of a Gram matrix: $L = D^TD$, for some matrix $D$. 
By further decomposing the columns of $D$ as the product of a scalar term $g$ and a normalized feature vector $\phi$, the matrix $L$ can be expressed as $L_{ij} = g_i\phi_i^{\top}\phi_j g_j,$ where $\phi$ is the normalized feature vector with $\phi_i^\top\phi_i = 1$, and $\phi_i^\top\phi_j \leq  \phi_i^\top\phi_i, \forall i,j \in \mathcal{Y} $. 
Therefore, for a single item $\mathbf{Y} = \{i\}$, the probability $\mathcal{P}( \mathbf{Y} = \{i\}) \propto g_i \phi_i^{\top}\phi_i g_i = g_i^2$ which indicates that a higher value $g_i$ increases the probability of selecting item $i$.
Thus, $g$ serves as a reasonable measure of quality.
For $\mathbf{Y} = \{i,j\}$, the probability 
\begin{align}
    \mathcal{P}( \mathbf{Y} = \{i,j\}) \propto L_{ii}L_{jj}-L_{ij}L_{ji} = g_i^2g_j^2\!\left( 1-\phi_i^{\top}\phi_j\phi_j^{\top}\phi_i \right)\notag,
\end{align}
where the inner product of feature vectors $\phi_i^{\top}\phi_j$ quantifies the similarity between items $i$ and $j$.
A higher inner product indicates a larger similarity, thereby reducing the likelihood of selecting these two items simultaneously. 
We refer to the similarity measurement defined by $\phi$ as the { \it cosine similarity}.
For learning purposes, it is often beneficial to decouple the quality and similarity model by defining the similarity value $S_{ij}= \phi_i^{\top}\phi_j$
with the corresponding similarity matrix $S_Y=[s_{ij}]_{i,j\in Y}$. 
Consequently, the matrix $L$ can be expressed as $L_{ij} = g_i S_{ij} g_j$. 
Then, the probability assigned by a DPP to the subset $Y$ becomes
 \begin{align}
     \mathcal{P}_L(Y) \propto \det(L_Y) = \left(\prod_{i \in Y} g_i ^2\right) \cdot \det(S_Y).
     \label{eq::DPP::S_ij::p_l}
 \end{align}
Drawing from this understanding, several important properties of DPP can be verified.
In particular, items with lower similarity (i.e., more orthogonal feature vectors) are preferable because they increase the volume of the parallelepiped spanned by them. 
\begin{figure*}[hbpt]
\centering
\includegraphics[width=\textwidth]{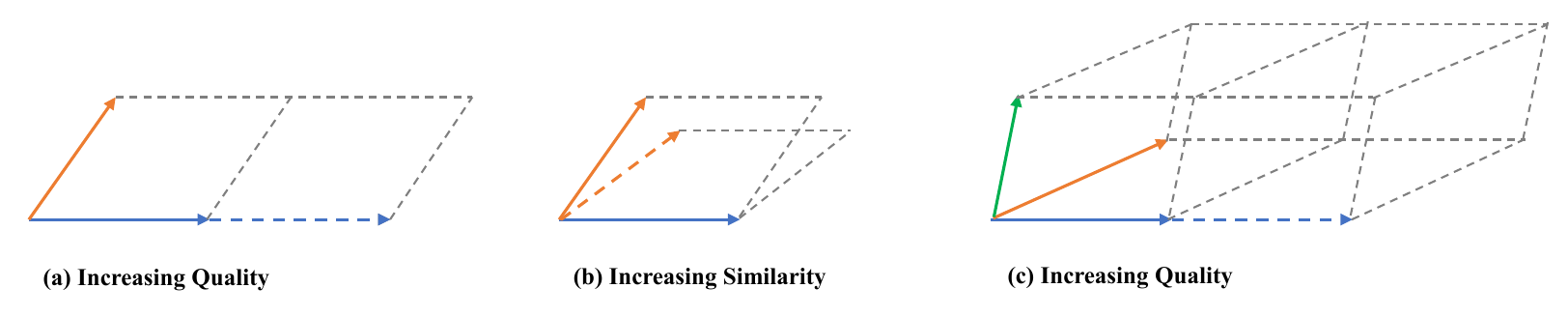}
\caption{ In DPP, the probability of occurrence of a set $Y$ depends on the volume of the parallelopiped with sides $g_i$ and angles proportional to $S_{i,j}$:  (a) Increasing the quality of one item, (b) decreasing the similarity of times, (c) increasing the quality of one item. }
\label{geometric} 
\end{figure*} 
Consequently, items with parallel feature vectors are unlikely to be chosen simultaneously since their feature vectors define a degenerate parallelepiped, yielding a selection probability of zero. 
Moreover, when all else is equal, items with large quality values $g$ are more likely to be chosen because they multiply the spanned volumes of subsets containing them. 
A visual representation of these properties is provided in~\figref{geometric}.
Focusing solely on quality can lead to a repetitive selection of similar high-quality items, while considering only the similarity might exclude the most important items, focusing instead on low-quality outliers.  
{\em DPPs balance quality and similarity by preferring to choose high-quality and low-similarity items simultaneously. }

 \subsection{Conditional DPPs} \label{subsec:ConditionalDPP}
Many learning applications are primarily driven by input data.
For example, in a search engine, users need to input keywords to display recommended results. 
Conditional DPPs are required to model these input-driven problems.
To develop the conditional DPP, let $X$ represent an external input set 
and $\mathcal{Y}(X)$ denote the collection of all possible subsets associated with that input $X$. 
Then, the matrix $L(X)$ can be represented as $L_{ij}(X) = g_i( X) S_{ij}(X) g_j(X)$, where $g_i(X) $ and $S_{ij}(X)$ denote the quality and similarity of the input $X$, respectively. 
More generally, the conditional DPP assigns probability to $Y \subseteq \mathcal{Y}(X)$ that is parameterized in terms of a generic $\boldsymbol{\theta}$ as 
\begin{equation}
	\mathcal{P}_{\boldsymbol{\theta}}(\mathbf{Y} = Y \mid X)  =  \frac{\det (L_Y(X;\boldsymbol{\theta}))}{\det (L(X;\boldsymbol{\theta})+I)}.
 \label{eq::conditionalDPP::p_theta}
\end{equation}
Now, assume we have a sequence of data samples $\left\{\left( X^{(t)}, Y^{(t)}\right)\right\}_{t=1}^{T}$, which are drawn independently  for a distribution over pairs $\left(X, Y\right)\in \mathcal{X} \times 2^{\mathcal{Y}(X)}$, where $\mathcal{X}$ is an input space and $\mathcal{Y}(X)$ is the associated ground set for input $X$. 
The objective now is to learn the optimal $\boldsymbol{\theta}$ from the input data. 
We achieve this by using the maximum likelihood estimate (MLE), which involves maximizing the conditional log-likelihood function.
Formally, the estimation/learning problem can be formalized as follows
 \begin{align}
 	\boldsymbol{\hat{\theta} }= \arg \max_{\boldsymbol\theta} \mathcal{L} (\boldsymbol{\theta};X),
  \label{eq::MLE::estimator}
 \end{align}
 where $\mathcal{L} (\boldsymbol{\theta}; X) $ is the conditional log-likelihood function defined as
 \begin{align}
 \mathcal{L} (\boldsymbol{\theta};X) 
 = \sum_{t = 1}^{T} \log \mathcal{P}_{\boldsymbol{\theta}}\!\left(Y^{(t)}\mid X^{(t)}\right).
  \label{eq::MLE::L_theta1}
 \end{align}
 Substituing~\eqref{eq::conditionalDPP::p_theta} into~\eqref{eq::MLE::L_theta1}, we have
 \begin{align}
 \mathcal{L} (\boldsymbol{\theta};X) 
 =& \sum_{t = 1}^{T} \bigg\{\log \det\!\left(L_{Y^{(t)}}\!\left(X^{(t)};\boldsymbol{\theta}\right)\right)\\
 &- \log \det\!\left(L\!\left(X^{(t)};\boldsymbol{\theta}\right)+{\rm I}\right) \bigg\}.
 \label{eq::MLE::L_theta2}
 \end{align}
If the gradient of objection function $\mathcal{L} (\boldsymbol{\theta}; X)$  exists and is computable,  standard algorithms such as gradient ascent or L-BFGS~\cite{nocedal1980updating} can be leveraged to find $\boldsymbol{\hat{\theta} }$. 

\subsection{Inference} \label{subsec:Inference}
The inference phase involves obtaining an estimated $\hat{Y}$ given an input $X$ using the trained conditional DPP.
In this section, we describe two methods for estimating $\hat{Y}$. 
\subsubsection{Sampling from DPP}
The first approach is to draw a random sample from the DPP, i.e., $\mathbf{Y} \sim \mathcal{P}_{\boldsymbol{\theta}^{*}} (\cdot|X) $ and set $\hat{Y} = \mathbf{Y}$. 
To achieve this, we first discuss the sampling scheme for a general DPP which naturally extends to the conditional DPP sampling. We begin by considering a special class of DPPs known as the elementary DPP and will use this method to draw samples from a general DPP. 
A DPP is called \textit{elementary} if every eigenvalue of its marginal kernel $K$ lies in $\left\{0,1\right\}$. 
Thus an elementary DPP can be denoted as $\mathcal{P}^{V}$ where $V = \left\{\mathbf{v}_1, \dots, \mathbf{v}_k\right\} $ is the set of $k$ orthonormal vectors such that $K^V = \sum_{\mathbf{v} \in V} \mathbf{v} \mathbf{v}^{\top}$. We now demonstrate that the samples drawn according to $\mathcal{P}^V$ always have fixed sizes. 
\begin{lemma} 
If $\mathbf{Y} \sim \mathcal{P}^V$, then $|\mathbf{Y}|=| V|$ almost surely.
\label{lemma1}
\end{lemma}
\begin{IEEEproof}
If $|Y| > |V|$, $\mathcal{P}^V(Y\subseteq \mathbf{Y})=0 $ since $\text{rank} (K^V) = |V|$. Hence $|\mathbf{Y}| \leq | V|$.
Now, we have $\mathbb{E}\!\left[|\mathbf{Y}|\right]=\mathbb{E}\!\left[\sum_{n=1}^{N}\mathbf{1}(a_n\in \mathbf{Y})\right]=\mathbb{E}\sum_{n=1}^{N}\!\left[ \mathbf{1}(a_n\in \mathbf{Y})\right] =\sum_{n=1}^{N}K_{n,n}=\text{trace}(K)=|V|.$
\end{IEEEproof}

We want to find a method to draw a sample $Y \subseteq \mathcal{Y}$ with length $k = |V|$.  
By Lemma~\ref{lemma1}, we have $\mathcal{P}^V(Y)=\mathcal{P}^V(Y\subseteq \mathbf{Y}) = \det \! \left(K_Y^V\right)$. 
Next, we present an iterative sampling scheme that samples $k$ elements from $\mathcal{Y}$ without replacement, ensuring that the joint probability of obtaining $Y$ is $\det(K_Y^V)$. 
Without loss of generality, we assume $Y = \{1,2,3,\dots,k\}$. 
Let $B = [\mathbf{v}_1^\top,\dots,\mathbf{v}_k^\top]^\top$ be the matrix whose rows are the eigenvectors of $V$. 
With this definition, we have $K^V=BB^\top$ and the determinant $\det\!\left( K_Y^V \right)=\left(\text{Vol}\!\left( \{\mathbf{b}_i\}_{i\in Y}\right)\right)^2$, where $\text{Vol}\!\left( \{\mathbf{b}_i\}_{i\in Y}\right)$ is the volume of the parallelepiped spanned by the column vectors $(\mathbf{b}_i \textit{-}\text{s})$ of $B$. 
Now, $\text{Vol}\!\left(\{ \{\mathbf{b}_i\}_{i\in Y}\}\right) = ||\mathbf{b}_1||\text{Vol}\!\left(  \{\mathbf{b}_i^{(1)}\}_{i=2}^k \right) $, where $ \mathbf{b}_i^{(1)}=\text{Proj}_{\perp b_1} \mathbf{b}_i $ denotes the projection of $ \{\mathbf{b}_i\}$ onto the subspace orthogonal to $\mathbf{b}_1$. 
Proceeding recursively, we obtain $\det\!\left(K_Y^V\right)=\left(\text{Vol}\!\left( \{\mathbf{b}_i\}_{i\in Y}\right)\right)^2=\!\left\| \mathbf{b}_1 \right\|^2 \times \!\left\| \mathbf{b}_2^{(1)} \right\|^2 \times\dots \times \!\left\| \mathbf{b}_k^{(1,\dots,k-1)} \right\|^2$. 
Thus, the $j^{th}$ step $(j>1)$ of the sampling scheme assuming $y_1=1,\dots,y_{j-1}=j-1$ is to select $y_j=j$ with probability proportional to $||\mathbf{b}_j^{(1,\dots,j-1)}||^2$ and project $\{ \mathbf{b}_j^{(1,\dots,j-1)} \}$  to the subspace orthogonal to $\mathbf{b}_j^{(1,\dots,j-1)} $. 
This iterative procedure guarantees that $\mathcal{P}^V(Y)=\det(K_Y^V)$. 

With this sampling scheme for an elementary DPP, we can now draw samples from a DPP. 
This scheme is enabled by the fact that a DPP can be expressed as a mixture of elementary DPPs. 
The following lemma formally states this result.

\begin{lemma} 
A DPP with kernel $L = \sum_{n=1}^N \lambda_n\mathbf{v}_n\mathbf{v}_n^\top$  is a mixture of elementary DPPs
\begin{equation}
    \mathcal{P}_L = \sum_{J\subseteq\{1,\dots,N\}} \mathcal{P}^{V_J} \prod_{n \in J} \frac{\lambda_n}{1+\lambda_n},
\label{eq::inference::mix_element}
\end{equation}
where $V^J=\{\mathbf{v}_n\}_{n\in J}$.
\label{lemma2}
\end{lemma}
\begin{proof}
Please refer to~\cite[Lemma 2.6]{Kulesza_2012} 
\end{proof}
Thus, given an eigendecomposition of $L$, the DPP sampling algorithm can be separated into two main steps: $(1)$ sample an elementary DPP $\mathcal{P}^{V_J}$ with probability proportional to $\prod_{n\in J}\lambda_n$, and $(2)$ sample a sequence of length $|J|$ from the elementary DPP $\mathcal{P}^{V_J}$. The steps discussed above are summarized in Alg.~\ref{alg1}.

\begin{algorithm}
\caption{Sampling form a DPP}\label{alg1}
\begin{algorithmic}[1]
\Procedure{SampleDPP}{$L$}
\State Eigen decomposition of $L$: $L =\sum_{n=1}^N\lambda_n\mathbf{v}_n\mathbf{v}_n^\top$ 
\State $J = \varnothing$
\For{\texttt{n = 1,\dots,N}}
    \State $J \gets J\cup \{n\}$ with probability $\frac{\lambda_n}{1+\lambda_n}$
    \State $V\gets\{ \mathbf{V_n}\}_{n\in J}$
    \State $Y \gets \varnothing $
    \State $B = [\mathbf{b}_1,\dots,\mathbf{b}_n] \gets V^\top$
\EndFor
\For{$1$ to $|V|$}
    \State  select $i$ from $\mathcal{Y}$ with probability $\propto ||\mathbf{b}_i||^2$
    \State  $Y \gets Y \cup \{i\}$
    \State  $\mathbf{b}_j \gets \text{Proj}_{\bot \mathbf{b}_i}\mathbf{b}_j$
\EndFor
\Return $Y$
\EndProcedure{}
\end{algorithmic}
\end{algorithm}

\subsubsection{MAP Inference} A more formal sampling approach is to obtain the maximum a posteriori (MAP) set, i.e., $\hat{Y} = \arg \max_{Y \subseteq \mathcal{Y}(X)} \mathcal{P}_{\boldsymbol{\theta}^{*}(Y|X)}=\arg \max_{Y \subseteq \mathcal{Y}(X)} \det (L_Y(X;\boldsymbol{\theta^*}))$. 
But, finding $\hat{Y}$ is NP-hard because of the exponential order search space $Y \subseteq \mathcal{Y} (X)$. 
However, we can construct a computationally efficient MAP inference algorithm using the submodularity of function $\log(\det(L_Y))$. 
Define the multilinear extension of $\log(\det(L_Y))$ as
\begin{align}
    F(\boldsymbol{x}) = \log\sum_Y\prod_{i\in Y} x_i \prod_{i\notin Y} (1-x_i) \det(L_Y),
\end{align}
where the $\boldsymbol{x}$ is a vector and $x_i$ represents the probability that set $Y$ containing the item $i$. 
For a PSD matrix $L$, we have 
\begin{align}
    F(\boldsymbol{x}) = \log\det(\text{diag}(\boldsymbol{x})(L-{\rm I})+{\rm I}),
\end{align}
and 
\begin{align}
    \frac{\partial}{\partial x_i}F(\boldsymbol{x}) = \text{tr}((\text{diag}(\boldsymbol{x})(L-{\rm I})+{\rm I})^{-1}(L-{\rm I})_i),
\end{align}
where $(L-{\rm I})_i$ denotes the matrix obtained by zeroing all except the $i$th row of $L-{\rm I}$.
Then, we can relax the discrete optimization problem into a continuous one to find the optimal vector $\boldsymbol{x}$ that maximizes the objective function $\det(L_Y)$. 
After obtaining the optional $\boldsymbol{x}$, we have the following theorem to round the non-integer $x_i$ with a predefined threshold $\delta$.
\begin{theorem}
    If $S =[0,1]^N$, then for any local optimum $\boldsymbol{x}$ of $F$, either $\boldsymbol{x}$ is integral or at least one fractional coordinate $x_i$ can be set to $0$ or $1$ without lowering the objective.
\end{theorem}
\begin{IEEEproof}
    Please refer to~\cite{gillenwater2012near} for the proof.
\end{IEEEproof}
The MAP sampling algorithm for DPP is given in Alg.~\ref{alg3} and Alg.~\ref{alg4}.

\begin{algorithm}
\caption{Local Optimal}\label{alg3}
\begin{algorithmic}[1]
\Procedure{Local-Opt}{$L,F,S$}
\State $\boldsymbol{x}=\boldsymbol{0}$ 
\While{\texttt{not converged}}
    \State $\boldsymbol{y} \gets  \arg\max_{y^{\prime}\in S} \nabla F(x)^{\top}y^{\prime}$ 
    \State $\alpha \gets \arg\max_{\alpha^{\prime}\in[0,1]} F(\alpha^{\prime}\boldsymbol{x}+(1-\alpha^{\prime})\boldsymbol{y})$
    \State $\boldsymbol{x} \gets \alpha \boldsymbol{x}+(1-\alpha)\boldsymbol{y} $
\EndWhile
\Return $\boldsymbol{x}$
\EndProcedure{}
\end{algorithmic}
\end{algorithm}

\begin{algorithm}
\caption{MAP sampling form a DPP}\label{alg4}
\begin{algorithmic}[1]
\Procedure{SampleDPP}{$L,F,S,\beta$}
\State $S = [0,1]^{N}$
\While{\texttt{not converged}}
    \State $\boldsymbol{x} \gets \textsc{Local-Opt}{ (L,F,S)}$
    \State $\boldsymbol{y} \gets \textsc{Local-Opt}{ (L,F,S\cap \boldsymbol{y}^{\prime}|\boldsymbol{y}^{\prime}\leq(1-\boldsymbol{x}))}$ 
\EndWhile
\If{ $F_{\boldsymbol{x}} \geq F_{\boldsymbol{y}}$} 
\State $Y=\{i \mid x_i>\delta\}$
\Else
\State $Y=\{i \mid  y_i>\delta\}$

\Return $Y$
\EndIf
\EndProcedure{}
\end{algorithmic}
\end{algorithm}
We also compare the near-optimal MAP inference scheme with the random sampling one for DPPs in our numerical simulations.   


\section{ Link Scheduling for Ad Hoc Network } \label{sec:case1}
In our first case study, we consider the subset selection problem in device-to-device (D2D) networks, which are an important class of ad hoc networks. 
Unlike infrastructure-based networks, such as cellular networks, there is a lack of tight coordination across proximate links, which leads to classical medium access challenges and consequently, self-interference across links. 
Since mutual interference is the dominant factor affecting these networks, it is especially critical to develop efficient strategies for interference mitigation, which inspired us to consider them as our first case study. 

In this section, we present our DPP-based link scheduling scheme for selecting the optimal set of active D2D links that maximize the sum-rate of the network. 
This case study will demonstrate the effectiveness of our proposed scheme when the optimal subset is selected from a ground set that is completely governed by a random process. 
In the next section, we will extend our analysis to a more complex 3D cellular network with directional antennas, where the cellular network is modeled using a deterministic model. 
This will offer evidence for the effectiveness of our proposed approach to a subset selection problem where the ground set is governed by an underlying deterministic model and has a much more complicated interference structure because of the directional antennas.

\subsection{System Model and Problem Statement} \label{subsec:case1system}

\begin{figure}[hbpt]
\centering
\includegraphics[width=0.43\textwidth]{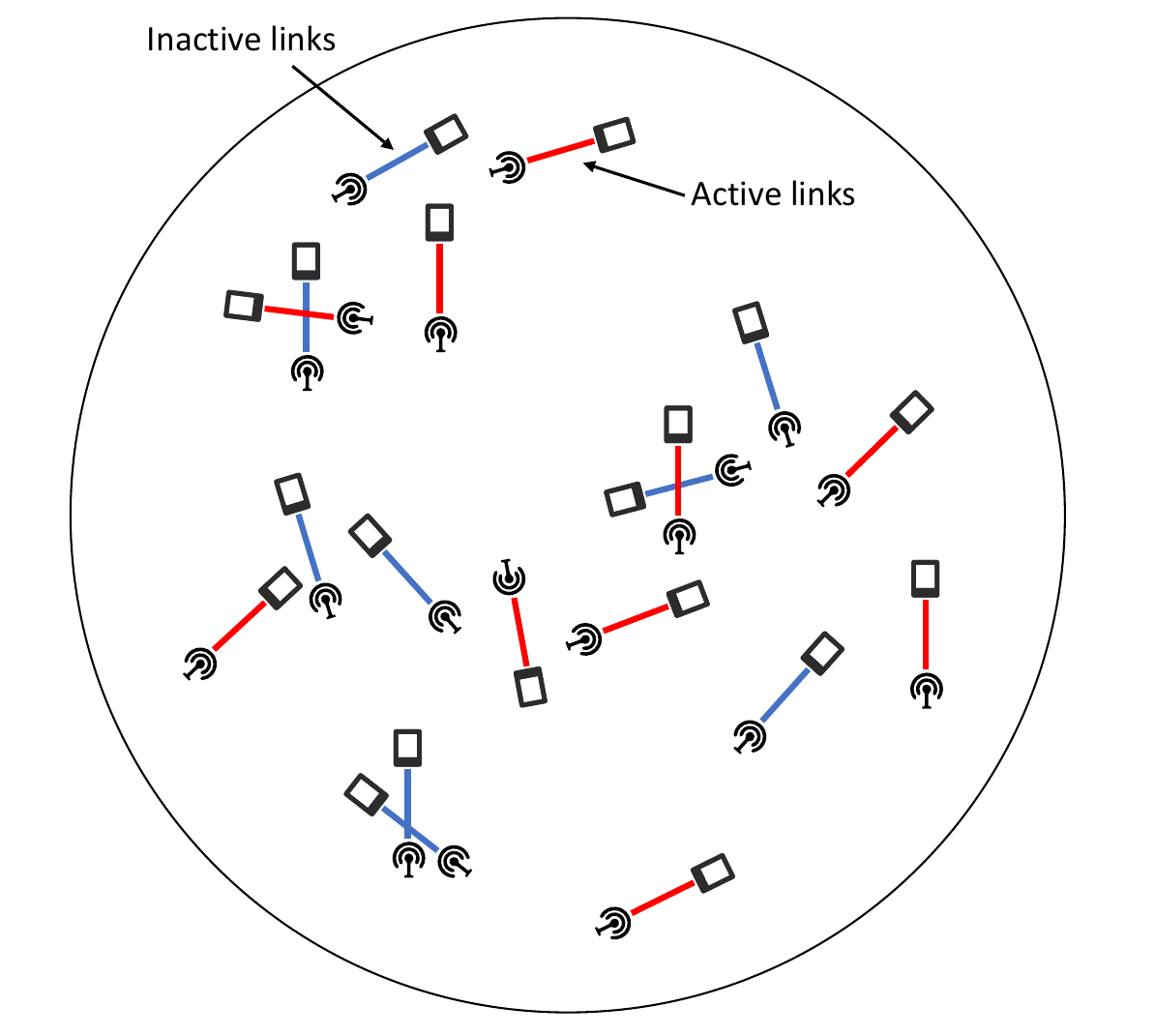} 
\caption{ Illustration of a network realization with the active link subset $\mathcal{E}^{*}$ for the ad hoc network. }
\label{adhoc::network}
\end{figure}

We consider a network with $M$ D2D links. 
The Txs locations of the D2D pairs are assumed to form a homogeneous PPP with density $\lambda_b$.
Each receiver (Rx) is distributed independently and uniformly at random with a fixed radius $d$ around its paired Tx. 
Both Tx and Rx are equipped with the omni-directional antennas. 
The network is modeled as a directed bipartite graph $\mathcal{G}: = \{\mathcal{N}_{t},\mathcal{N}_{r},\mathcal{E}\}$ where $\mathcal{N}_{t}$ and $\mathcal{N}_{r}$ represent the sets of vertices corresponding to the Txs and Rxs, respectively.
The set of directed edges is given by $\mathcal{E}:=\{e_i=(t_i,r_i): t_i\in \mathcal{N}_{t}, r_i\in \mathcal{N}_{r} \}$. 
Since each Tx has its dedicated Rx, the in-degree and out-degree of each node in $\mathcal{N}_{t}$ and $\mathcal{N}_{r}$ are one and we have $|\mathcal{N}_{t}|=|\mathcal{N}_{r}|=|\mathcal{E}|=M$. 
The illustration of the network topology is presented in~\figref{adhoc::network}.
Let $\mathcal{K}_{\mathcal{N}_{t},\mathcal{N}_{r}}^{\mathcal{W}}$ be the complete weighted bipartite graph on $\mathcal{N}_{t},\mathcal{N}_{r}$, where the weight $\mathcal{W}(j,i)=\zeta_{ji}$ represents the channel gain between Tx $j$ and Rx $i$ for all $j\in \mathcal{N}_{t}$, $i\in \mathcal{N}_{r}$. 
The downlink interference $I$ experienced by Rx $i$ is given by:  $I_{i} = \sum_{e_j\in \mathcal{E}}^{j \neq i} P_{j}\zeta_{ji},$ 
where $P_{j}$ is the downlink transmit power of Tx $j$  and $\zeta_{ji}= G_{ji} L_{ji}$, with $G_{ji}$ represents the antenna gain of Tx $j$ along the direction of Rx $i$ and $L_{ji}$ is the pathloss. 
For the simplicity of analysis, we utilize a distance-based pathloss, i.e.  $L_{ji} = d_{ji}^{-\beta}$, where $d_{ji}$ is the Euclidean distance between Tx $j$ and Rx $i$ and $\beta$ is the pathloss exponent. 
Extension to more general pathloss models is straightforward. 
Accordingly, the SINR of Rx $i$ is expressed as
\begin{equation}
    \gamma_{i} = \frac{P_{i}\zeta_{ii}}{I_i+N} = \frac{P_{i}\zeta_{ii}}{\sum_{e_j\in \mathcal{E}}^{j\neq i} P_{j}\zeta_{ji}+N},
     \label{eq::SINR::case1}
\end{equation}
where $N$ is the noise power. 
The data rate of the $i^{th}$ link is given by $\log_{2}\!\left(1+\gamma_{i}\right)$. 
We assume that each link is either active or inactive depending on whether its transmit power is at a high level $P_h$ or low level $0$. 
Our objective is to maximize the overall network sum-rate by selecting the subset of active links.
The sum-rate maximization problem can be defined as follows~\cite{weeraddana2012weighted}
\begin{subequations}
\begin{align}
    \max_{\{P_i\}} \quad  &\sum_{e_i\in \mathcal{E}}\log_2\!\left(1+\gamma_{i}\right),  \label{eq17a} \\
    \textrm{s.t.} \quad  &\gamma_{i} = \frac{P_{i}\zeta_{ii}}{\sum_{e_j\in \mathcal{E}}^{j\neq i} P_{j}\zeta_{ji}+N}, \label{eq17b}\\
     & P_{i}\in \{0,P_h\}, \label{eq17c} 
\end{align}
 \label{eq::opt::sumrate::case1}
\end{subequations}
\unskip\noindent where the downlink transmit power $\{P_{i}\}_{e_{i}\in\mathcal{E}}$ are optimization variables. 
The solution to this problem is the optimal set of active links denoted as $\mathcal{E}^{*}\subseteq \mathcal{E}$, such that $P_{i}=P_h, \forall \, e_{i}\in \mathcal{E}^{*}$ while $P_{i}=0, \forall \, e_{i} \in \mathcal{E} \backslash \mathcal{E}^{*}$. 
The optimization problem described in \eqref{eq::opt::sumrate::case1} is NP-hard. 
However, for bipartite networks, the original problem can be approximately solved using a low-complexity heuristic algorithm based on GP.

\subsection{Optimal Solution Based on GP} \label{subsec:case1gp}
Leveraging the monotonicity property of the $\log$ function, the sum-rate maximization problem in~\eqref{eq::opt::sumrate::case1} can be equivalently reformulated as
\begin{subequations}
\begin{align}
       \min \quad  & \prod_{e_{i}\in\mathcal{E}}(1+\gamma_{i})^{-1}, \label{eq19a} \\
    \textrm{s.t.} \quad \, & N \zeta_{ii}^{-1}P_{i}^{-1}\gamma_{i} + \sum_{j\neq i} \zeta_{ii}^{-1}P_{i}^{-1}P_{j}\zeta_{ji}\gamma_{i}\leq 1, \label{eq19b}\\
     & 0\leq P_{i}P^{-1} \leq 1, \label{eq19c}
\end{align}
\label{eq::opt::appro::sumrate::case1} 
\end{subequations} \unskip\noindent
where the constraint~\eqref{eq17c} is relaxed to the continuous interval $0\leq P_{i} \leq P$ and the equality constraint~\eqref{eq17b} is relaxed to inequality~\eqref{eq19b} since the objective function~\eqref{eq19a} is monotonically decreasing in $\gamma_l$. 
It can be verified that the constraints~\eqref{eq19b} and~\eqref{eq19c} are expressed in the standard form of GP problems~\cite{boyd2007tutorial}. 
However, the objective function is in fraction form, where both the denominator and numerator are polynomials.
This structure classifies the problem as a non-convex complementary GP, which is an NP-hard problem.
To solve this problem, we approximate the denominator of~\eqref{eq19a}, denoted as $f(\gamma_{i})$, using a monomial function $\bar{f}(\gamma_{i})$. 
Specifically, if function $\bar{f}(\gamma_{i})$ satisfies the following conditions for all $e_{i}\in\mathcal{E}$
\begin{subequations}
\begin{align}
        f(\gamma_l) & \geq \bar{f}(\gamma_l) ,\label{eq20a}\\
    f(\gamma_l^{\prime}) & = \bar{f}(\gamma_l^{\prime}) , \label{eq20b}\\
    \nabla f(\gamma_l^{\prime}) & =  \nabla \bar{f}(\gamma_l^{\prime}),
    \label{eq20c}
\end{align}
\label{eq::opt::appro::KT}
\end{subequations} 
\unskip\noindent where ${\gamma_l^{\prime}}$ is the solution in the previous iteration, then the solution $\gamma_{i}^{*}$ obtained from the approximated problem satisfies the Kuhn-Tucker (KT) conditions and is guaranteed to be a local minimum~\cite{inner1978}. 
In our approach, we choose the following monomial approximation
\begin{align}
    & \bar{f}(\gamma_i)  = \prod_{e_{i}\in\mathcal{E}}{ k_{i}\gamma_{i}^{\alpha_{i}}}.
    \label{eq::opt::appro::objec}
\end{align}
By enforcing condition~\eqref{eq20c}, we obtain $k_{i} = \frac{ (1+\gamma^{\prime}_{i})}{\gamma^{\prime \alpha_l}_{i}},$ and $\alpha_{i} = \frac{\gamma^{\prime}_{i}}{1+\gamma^{\prime}_{i}}.$
The next step is to verify inequality~\eqref{eq20a}. 
Substituting the expressions of $k_{i}$ and $\alpha_{i}$ into inequality~\eqref{eq20a} leads to
\begin{equation}
    \prod_{e_{i}\in\mathcal{E}}{(1+\gamma_i)}  \geq  \prod_{e_{i}\in\mathcal{E}}{\frac{ (1+\gamma^{\prime}_{i})} { {\gamma_{i}^{\prime \alpha_l} }} \gamma_{i}^{\alpha_{i}}},
    \label{eq::opt::appro::KT::inequ::1}
\end{equation}
for all $e_{i}\in\mathcal{E}$.
Then, we define  
 \begin{align}
     H(\gamma_{i}) = \left(\frac{ 1+\gamma^{\prime}_{i}}{1+\gamma_{i}}\right)  \!\left(\frac{ \gamma_{i}}{\gamma_{i}^{\prime}}\right)^{\frac{\gamma^{\prime}_{i}}{1+\gamma^{\prime}_{i}}},
 \end{align}
 and condition~\eqref{eq20a} requires $\log H(\gamma_i) \leq 0$.
By differentiating $\log H(\gamma_{i})$ with respect to $\gamma_{i}$, we obtain 
\begin{align}
   \frac{\partial \log H(\gamma_{i})}{\partial \gamma_{i}} &= \frac{\gamma^{\prime}_{i}}{\gamma_{i}(1+\gamma^{\prime}_{i})}-\frac{1}{1+\gamma_{i}} ,
   \label{eq::opt::appro::KT::derivation::1}\\
   \frac{\partial^2 \log H(\gamma_{i})}{\partial ^2 \gamma_{i}} &= - \frac{1}{\gamma^{\prime}_{i}(1+\gamma^{\prime}_{i})^2}.
   \label{eq::opt::appro::KT::derivation::2}
\end{align}
From equations \eqref{eq::opt::appro::KT::derivation::1} and \eqref{eq::opt::appro::KT::derivation::2}, it can be verified that $ \log H(\gamma_{i})$ is convex for $\gamma_{i} \geq 0$. 
By setting $ \frac{\partial \log H(\gamma_{i})}{\partial \gamma_{i}}=0$, we find that when $\gamma^{\prime}_{i} = \gamma_{i}$, the function $\log H(\gamma_{i})$ achieve its maximum value 0.
Therefore, it ensure that $ \log H(\gamma_{i}) \leq 0 $ for $\gamma_{i} \geq 0$ and satisfying constraint in~\eqref{eq20a}. 

Following this monomial approximation, we iteratively solve the GP approximation of the original optimization problem~\eqref{eq::opt::sumrate::case1}.
We use a predifined $\gamma^{\prime}$ to initialize the value of $k_{i}$ and $\alpha_{i}$. 
The approximate convex problem is then solved, and the solution from the current iteration is used to update the parameters $k_{i}$ and $\alpha_{i}$ for the next iteration.  
This iterative process repeats until the solution converges within a predetermined threshold. 
Then, we quantify the resulting optimal values $P_i$ to high power level $P_h$ and low power level $0$ with a predefined active threshold $P_{th}$.
The overall algorithmic steps are outlined in Alg.~\ref{alg2}. 

\begin{algorithm}
\caption{Optimization algorithm for problem (14)}\label{alg2}
\begin{algorithmic}[1]
\Procedure{SumRateMax}{$\mathcal{K}_{\mathcal{N}}^{W},\mathcal{E}$}
\State Initialization: given tolerance $\epsilon >0$, set $\mathbf{P}_0=\{P_{i,0}\}$. 
\State Set $n =1$. Compute $k_{n}$ and $\alpha_{n}$.
\Repeat 
    \State Solve the GP: 
    \begin{subequations}
    \begin{align}
        \min  \quad &\prod_{e_{i}\in\mathcal{E}}{k_i \gamma_{i}^{\alpha_i}}, \\
        \textrm{s.t.} \quad  & N \zeta_{ii}^{-1}P_{i}^{-1}\gamma_{i} + \sum_{j\neq i} \zeta_{ii}^{-1}P_{i}^{-1}P_{j}\zeta_{ji}\gamma_{i}\leq 1, \\
         & 0\leq P_{i}P^{-1} \leq 1, 
    \end{align}
    \label{eq::opt::alg}
    where $\{P_{i},\gamma_{i}\}_{e_{i}\in\mathcal{E}}$. 
    \end{subequations}
\State Denote the solution by $\{P_{i}^{*},\gamma_{i}^{*}\}_{e_{i}\in\mathcal{E}}$
\State n = n+1 
\Until $ \max_{e_{i}\in\mathcal{E}}|\gamma_{i}^{*}-\gamma_{i}^{(n-1)}|\leq \epsilon$
\If{ $P_{i}\geq P_{th}$} 
\State $P_{i}=P_h$
\Else
\State $P_{i}=0$
\EndIf
\Return $\mathcal{E}^{*}$
\EndProcedure{}
\end{algorithmic}
\end{algorithm}

\subsection{ Solution Based on DPPL} \label{subsec:case1DPP}
We implement the DPPL framework to solve the optimal subset selection problem. 
We first design the quality and similarity model and then define the matrix $L$. 
We train the DPP using a sequence of network realizations along with their optimal subset $X=\left(\mathcal{K}_{\mathcal{N}_{t}, \mathcal{N}_{r}}^{W},\mathcal{E},\mathcal{E}^{*}\right)$ that are obtained using Alg.~\ref{alg2}. 
 
\subsubsection{ Similarity  Model}\label{subsubsec:case1Q-Smodel}
The similarity matrix is constructed to capture the mutual interference among links.
Utilizing the popular Gaussian covariance function to construct PSD matrices, the similarity between Tx $i$ and Rx $j$ is defined as 
 \begin{equation}
  S_{ij}\!\left(X;\sigma\right) = \exp \!\left(-\frac{\|\mathbf{h}_{ij}(X)\|^2}{\sigma^2}\right),
  \label{eq::similarity::model::case1}
 \end{equation}
where $\mathbf{h}_{ij}(X)$ represents the feature that effectively capture the similarity between $i$ and $j$, and $\sigma$ is the learning parameter.
To minimize mutual interference between active links, we use mutual interference to define the similarity between links. 
Assuming omni-directional antennas and equal downlink transmit power for active links, interference can be characterized by a function of the inverse of the Euclidean distance $d_{ij}$ between the Tx $i$ and Rx $j$. 
Then, we propose a distance-based parametric similarity model where the similarity between Tx $i$ and Rx $j$ is given by the Gaussian covariance function of $d_{ij}$ and $d_{ji}$
 \begin{equation}
  S_{ij}\!\left(X;\sigma\right) = \exp\!\left(-\frac{d^2_{ij}+d^2_{ji}}{\sigma^2}\right),
  \label{eq::similarity::model::distance::case1}
 \end{equation}
with the property that $S_{ij}=S_{ji}$.
This formulation implies that active links exhibit some degree of distance repulsion to avoid mutual interference since a long distance $d_{ij}$ or $d_{ji}$ leads to a small similarity $S_{ij}$.
\subsubsection{ Quality Model}
In the quality model, links with higher received signal power should naturally be preferred while also accounting for the negative impact of interference.
Therefore, to capture these effects, the quality of links is parameterized as follows
 \begin{equation}
    g_i(X;\boldsymbol{\theta}) =\exp \!\left( \theta_1 \cdot P_{i}\zeta_{ii} + \theta_2 \cdot I_1 + \theta_3 \cdot I_2 \right),
\label{eq::quality::model::interfere::case1}
 \end{equation}
where $P_{i}\zeta_{ii}$ represents the received power at Rx $i$ and $I_1$ and $I_2$ denote the two strongest interference power values at Rx $i$. 

\subsubsection{ DPPL framework}\label{subsubsec:case1DPPL}
\begin{figure}[hbpt]
\centering
\includegraphics[width= 0.46\textwidth]{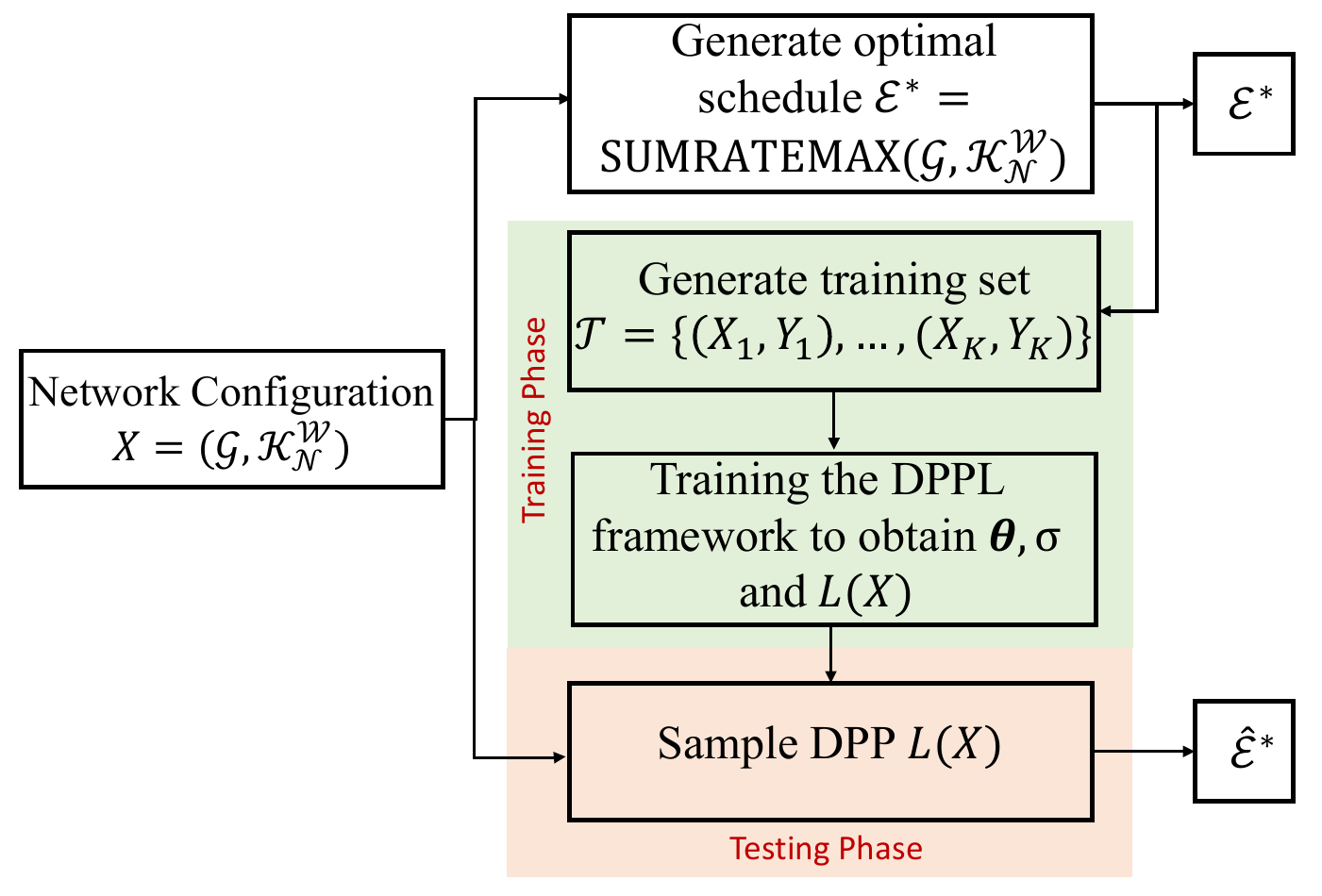} 
\caption{  Block diagram of the DPPL framework for the link scheduling problem. }
\label{diag}
\end{figure}

We now implement the proposed DPPL framework to solve this optimal subset selection problem.  
First, we use GP to obtain an approximate solution of~\eqref{eq::opt::sumrate::case1}.
Then, the resulting optimal values are quantified with threshold $P_{th}$ as described in Alg.~\ref{alg2}. 
The sequences of network realizations and their optimal subsets obtained by GP serve as the training dataset to obtain the optimal $\boldsymbol{\theta}$. 
Let $X_k = \left( \mathcal{K}_{\mathcal{N}_{t},\mathcal{N}_{r}}^\mathcal{W},\mathcal{E},\mathcal{E}^{*} \right)_k$ denote the $k^{th}$ realization of the network and its optimal subset.
The ground set for the DPPL is represented as $\mathcal{Y}(X) = \mathcal{E} $. 
During the testing phase, the subset sampled by DPPL is denoted by $\hat{\mathcal{E}^*}$. 
The block diagram of the DPPL framework is illustrated in~\figref{diag}. 

\subsection{Results and Discussions } \label{subsec:case1result}

\begin{figure}[hpt]
   \begin{minipage}{0.45\textwidth}
     \centering
     \includegraphics[width=0.95\linewidth]{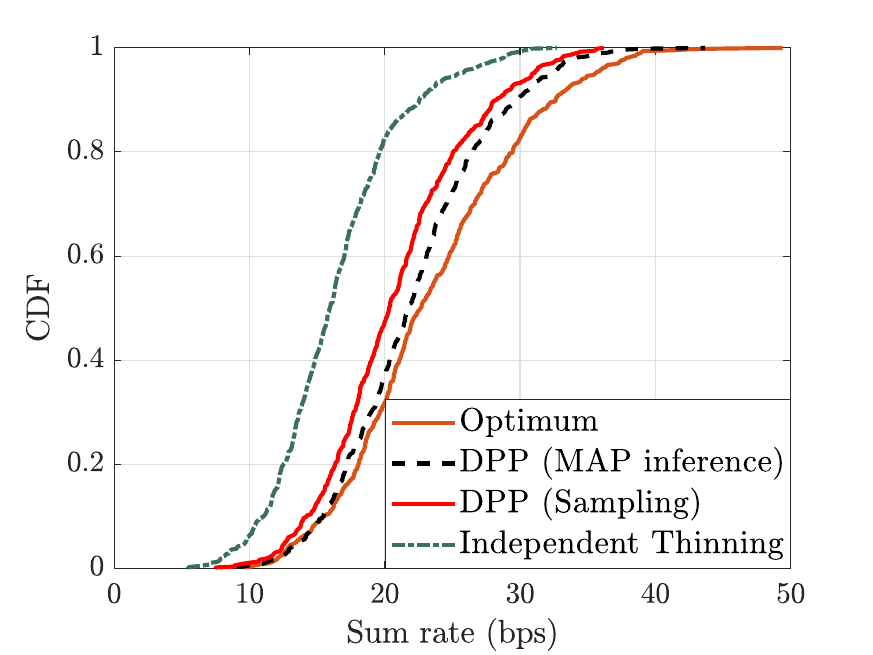} 
     \caption{The CDF of sum-rate obtained by different subset selection schemes including GP-based, DPPL and independent thinning in ad hoc network.}
     \label{fig:case1sumrate}
   \end{minipage}
   \begin{minipage}{0.45\textwidth}
     \centering
     \includegraphics[width=0.95\linewidth]{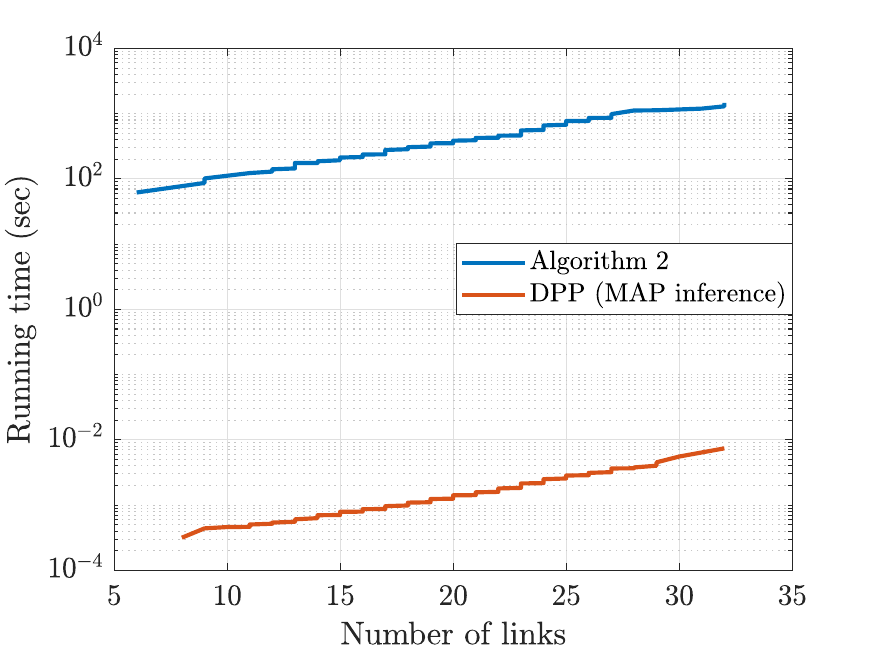}
     \caption{The comparison of running time of Alg.~\ref{alg2} and DPP in the testing phase in ad hoc network.}
     \label{fig:case1runningtime}
   \end{minipage}
\end{figure}

In this section, we demonstrate the performance of the proposed DPPL framework using numerical simulations. 
The network is constructed with $M$ Tx-Rx link pairs where $M \sim \text{Possion}\!\left(\lambda\right)$ and $\lambda = 20$. 
The channel gain is assumed to be dominated by the power law pathloss, i.e., $\zeta_{ij} = d_{ij}^{-\beta}$, where $\beta=2$ represents the pathloss exponent. 
Additionally, we set the high power level $P_h = 16\,$dBm, the lower power level $P_l = 0$ and the link active threshold $P_{th}= 3\,$dBm.
The training set $\mathcal{T}$ was generated by $ n= 300$ independent network realizations and their optimal subsets.
For comparison, we also evaluate a well-known SG-based method where the active links are modeled as independent thinning of the network~\cite{blaszczyszyn2019determinantal}. 
In this scheme, each link is assigned the high power $P_h$ according to an independent and identically distributed (i.i.d.) Bernoulli random variable with activation probability $p_a$. 
The $p_a$ is estimated from the data by averaging the activation of a randomly selected link, i.e., $p_a = \sum^W_{w=1} \mathbbm{1}(e_i\in \mathcal{E}_w^* )/W $. 
~\figref{fig:case1sumrate} shows the empirical cumulative distribution functions (CDFs) of the sum-rate achieved by different subset selection schemes.
Comparing these results obtained using GP, DPPL, and independent thinning shows that the DPPL framework closely approximates the maximum sum-rate.
Additionally, the MAP inference of the DPP provides a better estimation than DPP sampling. 
Furthermore, the sum-rate achieved by independent thinning is significantly lower than that obtained by using DPPL, which is not unexpected since the independent thinning scheme does not capture spatial repulsion across the links in $\mathcal{E}^*$. 

A key advantage of the proposed DPPL framework is its computational efficiency during the testing phase, as demonstrated in~\figref{fig:case1runningtime}. 
This result compares the running time of Alg.~\ref{alg2} with DPPL across different network sizes. 
The absolute values of running time were obtained by averaging over 200 iterations in the same computational environment. 
The exact details of the computational environment are immaterial in this representative comparison.
Note that DPPL not only obtains a near-optimal solution to this sum-rate maximization problem but is also significantly faster than the traditional optimization algorithms for estimating the optimal subset of links $\mathcal{E}^*$. 
Moreover, as the network size $M$ increases, solving the problem using Alg.~\ref{alg2} becomes infeasible due to its prohibitively long running time. 
This further emphasizes the scalability and efficiency of the DPPL framework which makes it a good solution for solving subset selection problems in large-scale networks. 

\section{ Link Scheduling for Drone Cellular Network} \label{sec:case2}
Cellular networks are typically interference-limited where managing interference is critical to improving SINR and user capacity. 
Directional transmission and link scheduling are used to reduce both inter- and intra-cell interference by orthogonalizing transmissions across space, time, and frequency dimensions.
We extend our analysis to a canonical cellular network setting. 
This generalizes our study from the previous section in three important ways. 
First, instead of the transmitters being modeled as a PPP, we consider a deterministic hexagonal model for the cellular network to demonstrate that DPPs can effectively model dependent thinning in such scenarios as well. 
Second, we assume directional transmissions, which introduce additional complexity to the interference structure, rendering traditional metrics, such as Euclidean distance, insufficient for defining similarity measures. 
This challenge motivated the exploration of alternative generative methods for constructing valid similarity matrices that accurately capture mutual interference in this practical network setting. 
Third, we assume that the BSs serve drones in 3D space, which means that the DPPs will now have to learn the structure of dependent thinning in the 3D space. 

\subsection{System Model}\label{subsec:case2system}

\begin{figure}[!htb]
   \begin{minipage}{0.5\textwidth}
     \centering
     \includegraphics[width=1\linewidth]{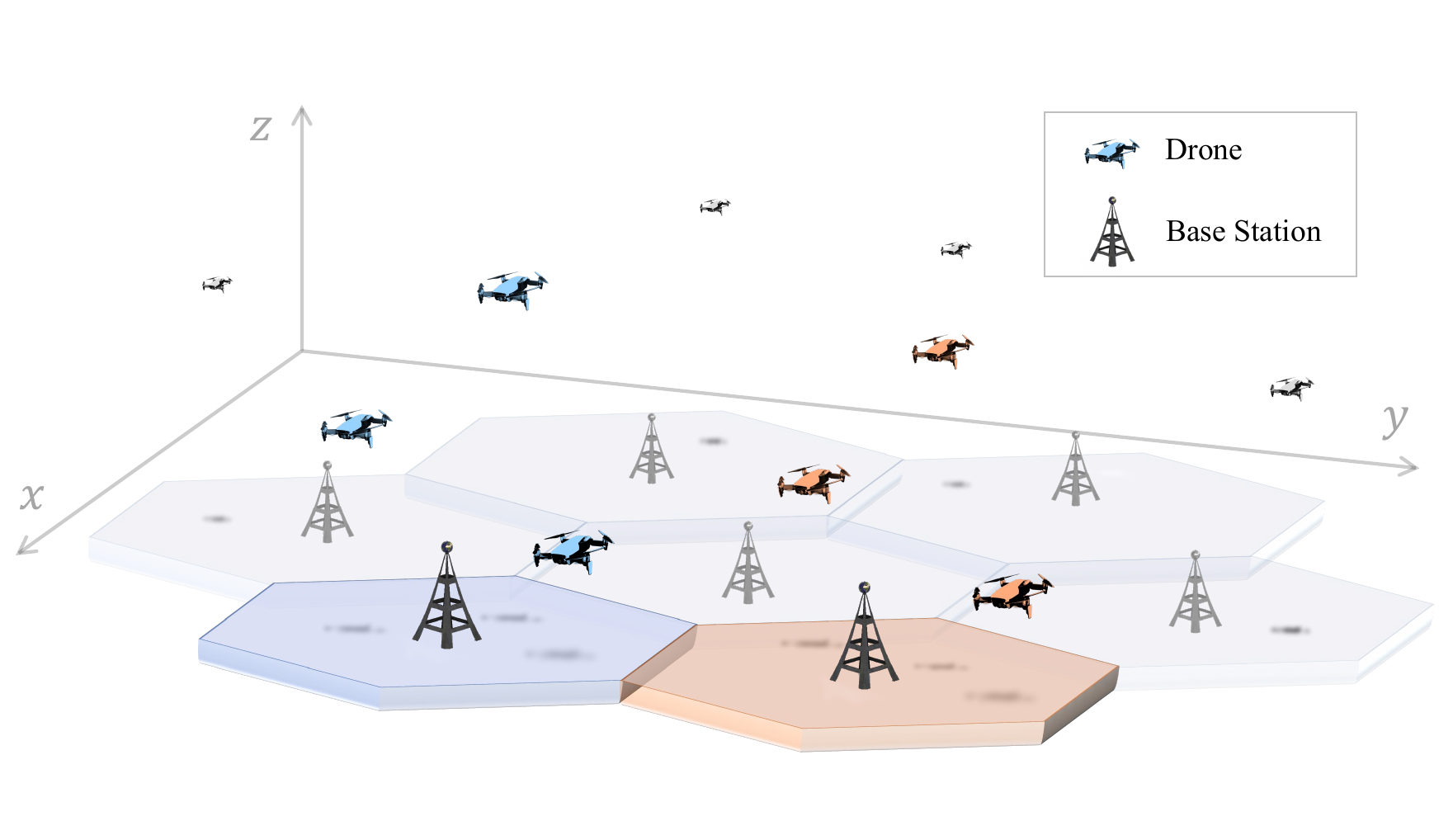} 
     \caption{ Illustration of the cellular network with seven cells. Only two of the seven cells are color-coded to highlight the spatial layout of the drones scheduled in each sector.}\label{fig::system::model::case2}
   \end{minipage}
   \begin{minipage}{0.5\textwidth}
     \centering
     \includegraphics[width=0.9\linewidth]{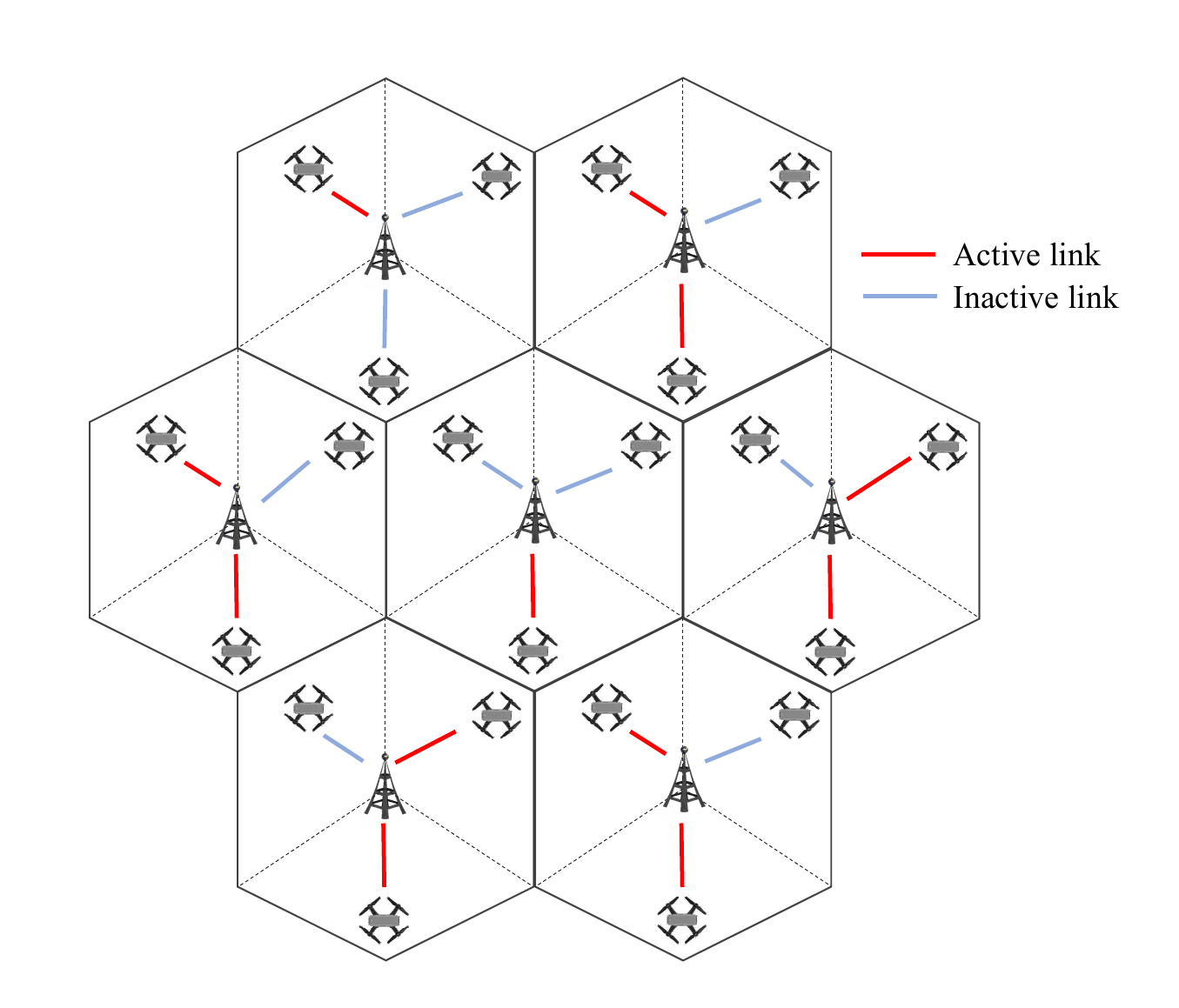}
     \caption{ The top view of a realization of the network with the optimal subset of BS-drone links $\mathcal{E}^*$. }\label{fig::system::relization} 
   \end{minipage} 
\end{figure}

First, we describe our system model for a cellular network that incorporates drones as user equipments (UEs).  
The BS sites are arranged in a 19-cell hexagonal grid with an inter-site distance (ISD) of $500\,$m. 
Each hexagonal cell is divided into three sectors, with each sector served by a BS. 
In each cell, the three BSs are co-located at the center with their beam azimuths offset by $120^\circ$.  
The BS antennas are downtilted by $100^\circ$ and are located at the height of $H_B=25\,$m. 
Drones are distributed uniformly and independently at random within the hexagonal grid and their heights are uniformly at random distributed between $H_l=1.5\,$m and $H_h=300\,$m above the ground. 
Each drone is equipped with an omni-directional antenna. 
After deploying the drones, we perform radio-distance-based association, where each drone is associated with the sector that offers the highest received power. 
Table~\ref{table1} lists the details of simulation parameters and assumptions, and~\figref{fig::system::model::case2} illustrates the drone cellular network configured with seven three-sector cells.
\begin{table*}[h]
\centering
\caption{Simulation Parameters}
\label{table1}
\begin{tabular}{|l|p{0.7\textwidth}|}
\hline
       Cell Layout            & Hexagonal grid, 19 cells, 3 sectors per Cell (ISD = $500\,$m)   \\
\hline
       Carrier Frequency      &   $6\,$GHz   \\
\hline
      System Bandwidth        &   $10\,$MHz  \\
\hline
       Height of BS    & $H_B=25\,$m   \\
\hline
      Noise Figure            & $7\,$dB  \\
\hline
      BS Tx Power             & $46\,$dBm   \\
\hline
      BS antenna pattern      & Antenna element pattern according to TR38.901. Vertical virtualization performed with downtilt angle $100^\circ$  \\
\hline
      Location of Drones     & Uniformly and independently at random distributed within its sector   \\
\hline
      Height of Drones       & Uniformly distributed between $H_l=1.5\,$m and $H_h=300\,$m  \\
\hline
      Antenna Pattern of Drones  &   Omni-directional\\
\hline
      Wrap-around  &   (Radio) Geometric distance based \\
\hline
\end{tabular} 
\end{table*}
Without loss of generality, we assume that only one drone is active per sector on a given time-frequency resource, with the active drone chosen uniformly at random from those associated with that sector. 
However, even with one active BS-drone link per sector, significant interference may arise due to the line-of-sight (LOS) propagation environment. 
Our objective is to implement the DPPL-based framework that selects an optimal subset of active BS-drone links to maximize the overall sum-rate.  

Let $\Phi_{\rm b}$ represent the set of BSs and $\Phi_{\rm u}$ represent the set of drones eligible for scheduling on a given time-frequency resource, where $|\Phi_{\rm b}|= |\Phi_{\rm u}|$.
The connectivity of the cellular network described above can be modeled as the directed bipartite graph $\mathcal{G}:=\{\Phi_{\rm b},\Phi_{\rm u},\mathcal{E}\}$ where $\mathcal{E}:=\{e_i=(t_i,r_i): t_i\in\Phi_{\rm b}, r_i\in \Phi_{\rm u} \}$ is the set of BS-drone pairs based on the association.
Our goal is to find the optimal subset of active BS-drone links denoted as $\mathcal{E}^*\subseteq \mathcal{E}$ to maximize the sum-rate of the network. The optimization problem is then formulated as follows
\begin{subequations}
\begin{align}
    \max_{\{P_i\}} \quad &\sum_{e_i\in \mathcal{E}}\log_2(1+\gamma_{i}), \\
    \text{subject to} \quad &\gamma_{i} = \frac{P_{i}\zeta_{ii}}{\sum_{e_j\in \mathcal{E}}^{j\neq i} P_{j}\zeta_{ji}+N}, \\
     & P_{i}\in \{0,P_h\}, 
\end{align}
\label{eq::opt::sumrate::case2}  
\end{subequations}
\unskip\noindent where $ i\in\Phi_{\rm b}$ is the index of BS, $j\in \Phi_{\rm u}$ is the index of drone. $P_i$ is the downlink transmit power of BS $i$ and the channel gain is given by $\zeta_{ij} = G_{ij}L_{ij}$, with $G_{ij}$ representing the antenna gain of $i^{th}$ BS along the direction of the $j^{th}$ drone. 
$L_{ij}$ is the pathloss between BS $i$ and drone $j$ and $N$ is the noise power.~\figref{fig::system::relization} shows a network realization with the optimal subset of BS-drone links $\mathcal{E}^*$. 
\subsection{Optimal Solution Based on GP} \label{subsec:case2gp} 
As demonstrated in Section~\ref{sec:case1}, the sum-rate maximization problem formulated above is NP-hard. 
To tackle this challenge, we employ a conventional GP-based method that approximates problem~\eqref{eq::opt::sumrate::case2} as a convex optimization problem. 
The approach described in Section~\ref{subsec:case1gp} and Alg.~\ref{alg2} to obtain optimal solution based on GP in the previous case is directly applicable to this case as well. Please refer to Section~\ref{subsec:case1gp} for more details. 
\subsection{Estimated Solution Based on DPPL} \label{subsec:case2DPP}
For link scheduling problems, the similarity matrix is designed to capture the mutual interference among BS-drone links.  
In our earlier analysis in Section~\ref{subsubsec:case1Q-Smodel}, under the assumption of omni-directional antennas for both Tx and Rx, we were able to simplify the construction of similarity matrix by expressing it as a direct function of the inverse of the Euclidean distance between the Tx and Rx (refer to equation~\eqref{eq::similarity::model::distance::case1}). 
However, the introduction of directional antennas complicates the interference structure, rendering Euclidean distance insufficient to capture mutual interference.
For example, BSs and users might be physically close but experience very low interference if the user is located around the null of the BS beam. 
Additionally, asymmetry in similarity arises due to directional antennas, where the interference caused by BS $j$ to Tx $i$ may differ from that caused by BS $i$ to Tx $j$.
Therefore, an alternative method for constructing asymmetric PSD similarity matrices is required to accurately capture these effects.

\subsubsection{Similarity and Quality Model} \label{subsubsec:case2Q-Smodel}
\begin{figure}[hbpt]
\centering
\includegraphics[width=0.48\textwidth]{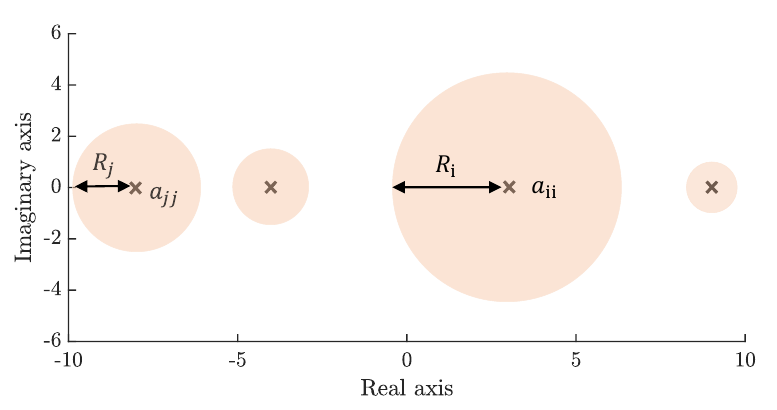} 
\caption{ Illustration of the Gershgorin discs $D$ with center $a_{ii}$ and radius $R_i(A), i\in\{ 1,2,...,n\}$ in yellow, which are derived for the eigenvalues of matrix $A$: every eigenvalue of matrix $A$ lies within at least one of the Gershgorin discs.}
\label{Ger}
\end{figure}

To maximize the overall network sum-rate, our primary objective is minimizing mutual interference among active BS-drone links. 
Motivated by this, we propose the interference-based parametric similarity model.
Defining the interference caused by BS $i$ to drone $j$ as $I_{ij}=P_{i}\zeta_{ij}$, the similarity model is defined as 
 \begin{equation}
  S\!\left(X;\sigma\right) = \sigma S\!\left(X\right),
  \label{eq::similarity::model::case2}
 \end{equation}
where $\sigma \in \mathbb R^+$ is the learning parameter and $S(X)$ is a matrix constructed based on $I_{ij}$.
Specifically, we define the elements of similarity $S$ as
\begin{align}
\label{eq::S::matrix::construction}
    S_{ij} = 
    \begin{cases}
    \max \{R_i(S)\}, & i = j \\
    I_{ij}, & i \neq j 
    \end{cases}, \, 
    \forall i\in\mathcal{N}_{t}, j\in\mathcal{N}_{r},
\end{align}
where $R_i(S) = \sum_{j\neq i}|S_{ij}|$.
Recall that the probability assigned by DPP is proportional to the determinant of $S(X)$ as denoted in~\eqref{eq::DPP::S_ij::p_l}.
Therefore, when incorporating directional antenna patterns, it is necessary to rigorously ensure that the matrix $S(X)$ is PSD.
To demonstrate this, we first employ the Gershgorin Circle Theorem to relate the entries of a matrix to its eigenvalues.  
\begin{theorem}
    Let $ A = [a_{ij}]\in \boldsymbol{M}_n $, where $\boldsymbol{M}_n$ represents the set of $n\times n$ matrices. 
    Define the absolute row sums of each row $i$ as $R_i(A) = \sum_{j\neq i}| a_{ij} |, i = 1,2,...,n$.
    Then, the $i$-th Gershgorin disc is given by
    \begin{align}
        \left\{ z \in \mathbb{C}: |z-a_{ii}| \leq R_i(A) \right\}, \forall i \in \{ 1,...,n\}.
        \label{eq::similarity::pro::GerDiscs}
    \end{align}
    All eigenvalues of matrix $A$ are in the union of Gershgorin discs, represented as 
    \begin{align}
        G(A) = \bigcup_{i=1}^n \!\left\{ z \in \mathbb{C}: |z-a_{ii}| \leq R_i(A) \right\}.
        \label{eq::similarity::pro::GerEigen}
    \end{align}
\end{theorem}
\begin{IEEEproof}
 Please refer~\cite[Theorem 6.1.1]{horn2012matrix} for the proof. 
\end{IEEEproof}
The Gershgorin Circle Theorem states that every eigenvalue of matrix $A$ lies within at least one of the Gershgorin discs $D(a_{ii}, R_i(A))$, as illustrated in~\figref{Ger}.

For real matrices, Gershgorin discs will reduce to intervals $T_i=[a_{ii}-R_i, a_{ii}+R_i ]$.
A real matrix (such as $S$) is PSD if and only if all of its eigenvalues are non-negative. 
This condition can be guaranteed by ensuring that all intervals $T_i$ lie on the positive side of the axis, which can be represented by $a_{ii} \geq 0$ and $|a_{ii}| \geq R_i(A)$ for all $i\in\Phi_{\rm b}$. We state these conditions formally in the next Proposition.
 \begin{prop}
 \label{proposition1}
      A matrix $ A = [a_{ij}]\in \boldsymbol{M}_n $ is a positive semidefinite matrix if:
     \begin{enumerate}
         \item $ |a_{ii}| \geq R_i(A), \forall i \in\{ 1,2,...,n \}$,
         \item $ a_{ii} \geq 0 $.
     \end{enumerate}
 \end{prop}
Now, one can easily construct a PSD matrix $S$ using constraints from Proposition~\ref{proposition1}. Note that under this construction, $S$ neither needs to be symmetric nor decomposable. 
It is noteworthy that even though our similarity matrix was based on mutual interference in this work, Proposition~\ref{proposition1} is more general and can be applied to other settings in which the similarity might depend on other aspects or features according to specific problems. 

\subsubsection{ Quality Model}
In the quality model, links with higher SINR should naturally be preferred. 
To capture this, we parameterize the quality function based on $\rm{SINR}$ as follows
 \begin{align}
     g_i(X;\theta) = \left(\theta \cdot \gamma_i\right)^{\frac{1}{2}},
     \label{eq::quality::model::SINR::case2}
\end{align}
where, $\gamma_i$ is the SINR of Rx $i$. 

 \subsubsection{Training DPPL framework} 
Incorporating the newly defined quality and similarity model into the DPPL framework, the log-likelihood function from equation~\eqref{eq::MLE::L_theta1} can be expressed as follows
 \begin{align}
&\mathcal{L}\!\left(\boldsymbol{X};\boldsymbol{\theta},\sigma\right) 
  = \sum_{t = 1}^{T} \bigg\{\log \det\!\left(L_{Y^{(t)}}\!\left(X^{(t)};\boldsymbol{\theta},\sigma\right)\right) \notag\\
  &- \log \det\!\left(L\!\left(X^{(t)};\boldsymbol{\theta},\sigma\right)+{\rm I}\right) \bigg\} \notag\\
  =& \sum_{t = 1}^{T} \bigg\{\log\!\left[ \prod_{i\in Y^{(t)} } \theta \cdot \gamma_i \cdot \det\! \left(\sigma S_Y(x)\right)\right]   \notag \\
  & - \log \!\left[\sum_{Y^\prime \subseteq \mathcal{Y}(X)} \prod_{i\in Y^\prime } \theta \cdot \gamma_i \cdot \det\! \left(\sigma S_{Y^\prime}(x)\right) \right] \bigg\} \notag,
 \label{eq::loglikelihood::case2} 
 \end{align} 
where $S_Y=[s_{ij}]_{i,j\in Y}$.
It is evident that the gradient of $ \mathcal{L} (\boldsymbol{X};\boldsymbol{\theta}, \sigma)$ with respect to both $\theta$ and $\sigma$ exists.
We train the DPP with a sequence of network realization and their optimal subsets, denoted by $X_{k}=\left(\mathcal{K}_{\mathcal{N}_{t}, \mathcal{N}_{r}}^{W},\mathcal{E},\mathcal{E}^{*}\right)$. 
Please refer to the section~\ref{subsubsec:case1DPPL} and~\figref{diag} for a detailed description of the training procedure.

\subsection{Result and Discussion} \label{sec:NumResults}

\begin{figure}
   \begin{minipage}{0.45\textwidth}
     \centering
     \includegraphics[width=0.95\linewidth]{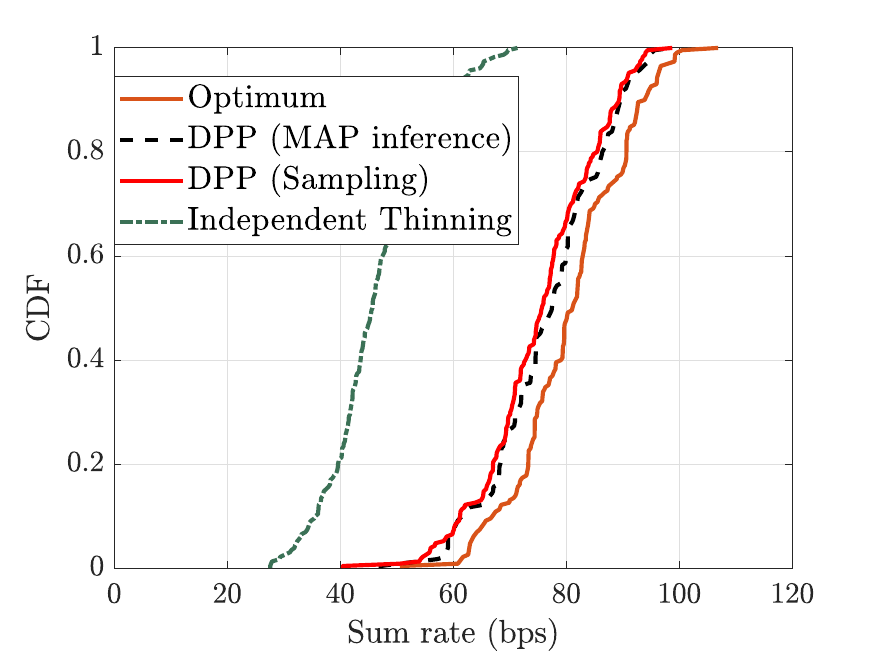} 
     \caption{The CDF of sum-rate obtained by different subset selection schemes including GP-based, DPPL and independent thinning in drone wireless network.}\label{fig:case2sumrate}
   \end{minipage}
   \begin{minipage}{0.45\textwidth}
     \centering
     \includegraphics[width=0.95\linewidth]{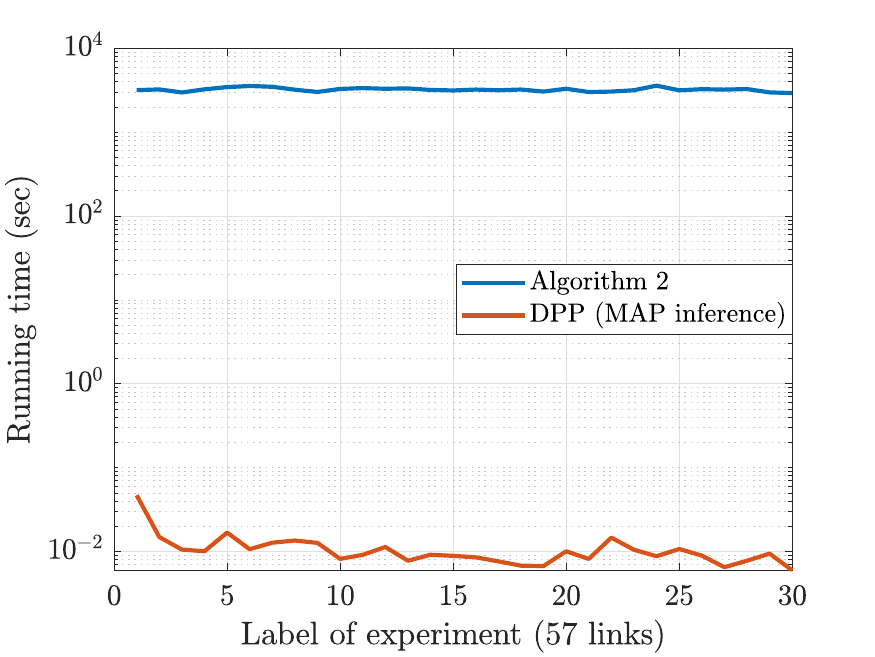}
     \caption{The comparison of running time of Alg.~\ref{alg2} and DPPL in the testing phase in drone wireless network.}\label{fig:case2runningtime}
   \end{minipage}
\end{figure}

We now demonstrate the performance of the DPPL framework with this new similarity model in~\eqref{eq::S::matrix::construction} through numerical simulations. 
The simulations are conducted on a cellular network comprising 19 cells, each divided into three sectors, with one BS acting as the transmitter in each sector. 
For each network realization, a large number $M$ of drones is generated and assigned to sectors using radio-distance-based association, where each drone associates with the sector with the highest received power. 
We set the high power level $p_h = 46\,$dBm, low power level $p_l= 0$, and the active threshold $p_{th}= 15\,$dBm.
The training set $\mathcal{T}$ is generated using $n = 200$ independent realizations of the network and their corresponding optimal subsets.~\figref{fig:case2sumrate} presents the empirical CDFs of the sum-rate achieved by different subset selection methods, including the GP solution obtained from Alg.~\ref{alg2}, DPPL, and independent thinning. 
Comparing these results shows that the DPPL closely approaches the optimal sum-rate. 

Furthermore, as was also the case in the previous case study, DPPL is significantly more efficient in terms of running time than the GP-based method. 
This is also supported by the results in~\figref{fig:case2runningtime}.
To compare the computational efficiency of DPPL and GP, we arbitrarily select $30$ realizations of the network and computed the optimal schedules using both approaches.  
The results show that the GP-based method is roughly $10^5$ times slower than solving it using DPPL. 
This is not surprising since GP {\em solves} the optimization problem, whereas DPPL simply obtains the optimal solution through efficient {\em sampling}.
This scalability makes the proposed DPPL particularly appealing for massive networks, such as large-scale IoT networks.


\section{Concluding Remarks}\label{sec:conclu}

In this paper, we introduced a DPP-based learning framework for solving general subset selection problems in large-scale wireless networks. 
After explaining the mathematical underpinnings of the proposed framework, we applied it to two canonical wireless network settings that cover a variety of scenarios of general interest. 
First, we applied it to the D2D network setting, which is an important subclass of ad hoc networks. 
We showed that the proposed DPPL is effective in solving the link scheduling problem by capturing the trade-off between signal strength and mutual interference among links to obtain the optimal set of active D2D pairs. 
While the base set in this example was completely governed by an underlying random process (point process), we extended this to a more complicated cellular network setting serving drones in the 3D space with directional antennas. 
This offers an example of a setup where the base set for subset selection is governed by a deterministic hexagonal cellular network model. 
Because of directional transmission, one could not rely on the traditional ways of constructing similarity matrices that unnecessarily impose decomposability and symmetry constraints on these matrices. 
Since these constraints do not hold in our second setting, we proposed a new method to generate these matrices using the Gershgorin Circle Theorem, which further enhances the applicability of the proposed DPPL framework to an even larger class of wireless setups with more complicated correlations. 

While this paper has presented a complete DPPL framework that is applicable to a variety of subset selection problems, one can think of many potential follow-up efforts. From the training perspective, it is important to incorporate adaptive and real-time learning mechanisms into the DPPL training phase to update the optimal model parameters. 
Approximation of optimal solutions using a DPP also opens up the possibility of deriving new performance bounds for these subset selection problems using tools from stochastic geometry. 
Finally, the framework can obviously be extended and applied to many new wireless scenarios, such as user group selection in downlink multi-antenna networks.

\bibliographystyle{IEEEtran}
\bibliography{hokie}

\end{document}